\documentclass[sn-basic]{sn-jnl}% Basic Springer Nature Reference Style/Chemistry Reference Style

\usepackage{graphicx}%
\usepackage{multirow}%
\usepackage{amsmath,amssymb,amsfonts}%
\usepackage{amsthm}%
\usepackage{mathrsfs}%
\usepackage[title]{appendix}%
\usepackage{xcolor}%
\usepackage{textcomp}%
\usepackage{manyfoot}%
\usepackage{booktabs}%
\usepackage{algorithm}%
\usepackage{algorithmicx}%
\usepackage{algpseudocode}%
\usepackage{listings}%
\usepackage[normalem]{ulem}
\theoremstyle{thmstyleone}%
\theoremstyle{thmstyletwo}%

\theoremstyle{thmstylethree}%

\begin{document}

\title[Article Title]{Controlling quantum phases with step-like electric potentials in one-dimensional Hubbard systems}


\author{\fnm{D.} \sur{Arisa}}

\author{\fnm{R. M.} \sur{Dos Santos}}

\author{\fnm{I. M.} \sur{Carvalho}}

\author*{\fnm{V. V.} \sur{França*}}\email{vivian.franca@unesp.br}

\affil{São Paulo State University (UNESP), Institute of Chemistry, 14800-090, Araraquara, São Paulo, Brazil}


\abstract{Quantum systems under electric {potentials} provide a powerful framework for uncovering and controlling novel quantum phases, especially in low-dimensional systems with strong correlations. In this work, we investigate quantum phase transitions induced by a {step-like} electric potential in a one-dimensional half-filled Hubbard chain. By analyzing {\it i)} tunneling {energy} and {local doublon response}, {\it ii)} charge and spin gaps, and {\it iii)} entanglement between the chain halves, we identify three distinct phases: Mott insulator, metal and band-like insulator. The metallic regime, characterized by the closing of both charge and spin gaps, is accompanied by an {electric-potential dependence of} kinetic energy and a quasi-periodic oscillatory behavior of {local doublon} response and entanglement. Although the metallic phase persists for different magnetizations, its extent in the phase diagram shrinks as spin polarization increases.}




\maketitle

\section{Introduction}\label{sec1}

A key feature of quantum systems is their pronounced sensitivity to external perturbations, such as electric {potentials}, as these enhance quantum fluctuations and lead to nontrivial emergent phenomena ~\cite{Goychuk01092005,RevModPhys.83.771}. In systems described by the Hubbard model, which balances electron motion and repulsion, complex phenomena such as superfluid phases~\cite{PhysRevA.86.033622,Canella2019,CANELLA2020123646,Arisa_2469_2020} and Mott-Anderson transitions~\cite{Canella_2024_2022,canellaVV} emerge.  {Electric potentials can strongly affect these systems, producing} undamped Bloch oscillations~\cite{ZHENG2021127112}, driving them to non-thermal steady states~\cite{PhysRevB.91.245153}, and even triggering transitions from metal to Mott insulator~\cite{PhysRevLett.114.226403, PhysRevB.109.115104}.

In one-dimensional (1D) systems, the reduced dimensionality amplifies quantum fluctuations, which significantly affect correlations and ordering~\cite{giamarchi2003quantum,Thierry_Giamarchi}. Also, the growing number of 1D materials~\cite{Jardine_2542_2021,Baydin_1535_2022,Laird_703_2015,Qu2022,Tang2023,Li01012012,Wang_791_2021,Li_1_2012,Alfieri_1521_2022} makes them especially relevant for exploring quantum phase transitions (QPTs). Although two-site (dimer) {Hubbard systems have long served as minimal models to explore the response to applied electrostatic potentials, including step-like profiles} analysis~\cite{BALCERZAK2017252, BALCERZAK20181069,VILLEGAS2024129898, HASEGAWA2005273, HASEGAWA20111486}, scaling up such systems introduces additional correlations~\cite{Manmana_1550_2004,Gerster_2016,PhysRevB.61.12496,Qu2022,PhysRevB.109.L100502,PhysRevB.106.195405,PhysRevB.110.115145,RevModPhys.60.1129,sanino2024,essler2005one}, arising from collective excitations, which cannot be captured by simple dimer model extrapolations.

{Beyond dimer analyses, theoretical studies of extended 1D systems have shown that self-consistent step-like electrostatic potentials can generate metallic phases at the interface between Mott and band insulators~\cite{naoto_95_2005,dagotto_74_2006}. In contrast, a linear Wannier–Stark potential produces interfaces between commensurate and incommensurate density regions that remain insulating, with a finite local density-of-states gap~\cite{nair_109_2024}. Thus, while self-consistent steps yield metallic interfaces and linear potentials preserve insulating behavior, the case of fixed step-like potentials without self-consistency remains unresolved. This question is particularly relevant to ultracold-atom experiments, where optical lattices enable precise engineering of external potentials~\cite{RevModPhys.80.885,doi:10.1126/science.aal3837,Schafer2020}, including state-dependent lattices capable of creating interfaces and step-like profiles~\cite{PhysRevLett.105.045303}.}

Here we investigate {quantum phase transitions induced by a static step-like electrostatic potential} in the many-site 1D Hubbard chains at half-filling. We {identify} three distinct phases: Mott insulator, metallic, and band insulator. {The metallic phase is} characterized by the vanishing of both charge and spin gaps, {and the charge gap decays linearly with increasing system size, in contrast to previous studies where metallicity was confined to a finite-width interface region~\cite{naoto_95_2005,dagotto_74_2006}. The metallic phase is also marked} by a field dependence of the kinetic energy and quasi-periodic oscillations in {the local doublon} response and entanglement. {These quantities are then used to characterize the metallic phase in magnetized systems, since the gap is only well defined at zero magnetization.} While the metallic phase {persists} across varying magnetizations, increased spin polarization progressively narrows its range in the phase diagram.

\section{Model and methods}

We consider a 1D Hubbard model under an electrostatic potential {polarized in opposite directions} ($+|V|$ and $-|V|$) on the two halves of the chain, resulting in a total potential difference of $2|V|$ between them. The corresponding Hamiltonian is 
\begin{eqnarray}
     H= \underbrace {-t\sum_{j=1}^{L-1} \sum_{\sigma} ({c_{j,\sigma}^{\dagger}c_{j+1,\sigma}} + c_{j+1,\sigma}^{\dagger}c_{j,\sigma})}_{H_t}
    + \underbrace{U\sum_{j=1}^{L}n_{j\uparrow}n_{j\downarrow}}_{H_U}
    + \underbrace{V\sum_{j=1}^{L/2}( n_{j}-n_{j+L/2})}_{H_V}, \label{terms_hamiltonian}
\end{eqnarray}
where the terms $H_t$, $H_U$, $H_V$ are associated with the kinetic, interaction ($U$) and electric potential energies, respectively. Here $L$ is the chain size, $c_{j\sigma}^{\left(\dagger\right)}$ is the electron annihilation (creation) operator, $\sigma=\uparrow,\downarrow$ labels the electron spin, and $n_{j}=n_{j\uparrow}+n_{j\downarrow}$ is the particle number operator at site $j$. We set the hopping amplitude $t$ as the energy unit.

At half-filling $n=1$ and zero net magnetization, $m=1/L\sum_{j}(n_{j,\uparrow}-n_{j,\downarrow}) = 0$, the Hamiltonian exhibits well-defined ground states in the interaction asymptotic limits. For $U\gg t \gg V$, the system manifests as a spin-density wave insulator {(i.e., a Mott insulator)}, with electrons occupying individual sites and equally likely to be in spin-up or spin-down states along the chain. For $U \ll t\ll V$, the lower-potential sites ($-|V|$) are doubly occupied by electrons with opposite spins, resulting in a state similar to a band insulator \cite{Feiguin_76_2007}, with both charge and spin degrees of freedom frozen. 

We are interested in the interplay between $U$ and $V$ in the intermediate regime. We thus analyze the average charge density in the lower-potential region, $n^{(-V)}=2/L\sum_{j=L/2}^{L} n_{j}$, the {doublon} response, via $\partial \Bar{w_2}/\partial V$, where $\Bar{w_2}$ is the average double occupancy, and the charge and spin gaps~\cite{Hohenadler2018}, 
\begin{eqnarray}
    \Delta_{c}&=&e_{0}(N-2,0)+e_{0}(N+2,0)-2e_{0}(N,0)\\
    \Delta_{s}&=&e_{0}(N,1)-e_{0}(N,0),
\end{eqnarray}
where $e_0(N,S^{z})$ represents the ground-state energy for a given number of electrons $N$ with total spin-z component $S^{z}$. In metallic phases, charge excitations occur without energy cost, thus $\Delta_{c}=0$; whereas in insulating phases, a finite charge gap  $\Delta_{c} > 0$ signals an energy barrier to charge transport. The energy associated with spin flips is quantified via the spin gap, therefore $\Delta_{s} = 0$ for gapless spin excitations, as in a typical spin-density wave or antiferromagnet phases \cite{EjimaEHM2007,PhysRevB.74.245110}; while $\Delta_{s}>0$ for spin-gapped phases, which may arise from dimerization or spin-singlet formation \cite{giamarchi2003quantum, essler2005one}. 

Since QPTs are driven by nonlocal quantum correlations~\cite{Islam2015}, entanglement has become a powerful tool for identifying and characterizing them~\cite{Pauletti_129824_2024,FRANCA201782, Canella_2024_2022,Arisa_2469_2020}. Thus we also analyze nonlocal correlations between the two halves of the chain via the entanglement entropy, $S=-\sum_{i}\lambda_{i}\log_{2}\lambda_{i}$, where $\left\{ \lambda_{i}\right\} $ are the eigenvalues of the reduced density matrix $\rho_{{\rm L/2}}={\rm Tr_{{\rm L/2}}\left|\Psi_0\right\rangle \!\left\langle \Psi_0\right|}$ for the ground state $\left|\Psi_0\right\rangle $~\cite{LAFLORENCIE20161,Witnessing_Entanglement,RevModPhys.80.517}. $\left|\Psi_0\right\rangle $ is obtained via density matrix renormalization group (DMRG) calculations~\cite{dmrg1,dmrg2}, implemented with the ITensor library~\cite{itensor}. {All DMRG simulations were performed conserving the total charge and total spin quantum numbers, with the initial state prepared within the corresponding sector. Typically, we start from a Néel state in the non-magnetized sector, while for studying magnetization we flip the boundary spins. Although different initial states can be chosen, as long as they respect the same quantum numbers, they converge to the same results for energy and correlations.} The precision was controlled by the bond dimension $D$, whose maximum is set at $\sim2^{11}$. We apply a warm-up stage in the DMRG sweeps, gradually increasing $D$ by $\sim 2^{3}$ per sweep. Calculations continued until the ground-state energy convergence to at least $10^{-7}$. In all cases we considered open boundary conditions for system sizes of up to $2^{5}$ sites, ensuring that our analysis captures the relevant physics while minimizing finite-size effects.

\section{Results}\label{sec2}

\begin{figure}[t!]  
    \centering   
    \includegraphics[scale=0.70]{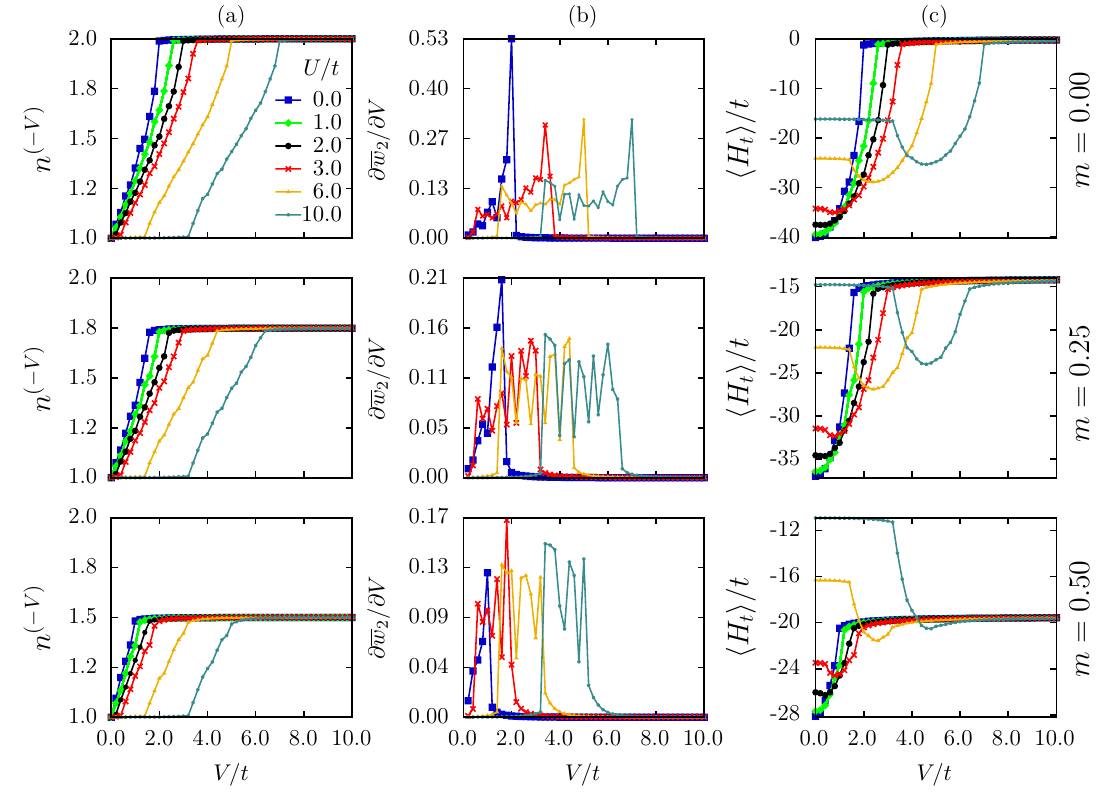}
    \caption{(a) Average charge density $n^{(-V)}$ in the lower-potential region, (b) {doublon} response $\partial \Bar{w_2}/\partial V$, and (c) kinetic energy $\langle H_{t} \rangle/t$  as a function of the electric potential $V$ for different interactions $U$ and magnetizations $m$. {In panels (b), some values of $U$ were omitted to improve the visualization, preserving the overall trend with interaction.}}
    \label{fig:occupation}
\end{figure}

\subsection{Tunelling and {doublon} mechanisms}

As we consider half-filled chains, for $V=0$ $n^{(-V)}=1$ as shown in Fig.~\ref{fig:occupation}(a). The electric {potential} shifts electrons toward lower-potential sites, $n^{(-V)}>1$, and enhances the {doublon} response, $\partial \Bar{w_2}/\partial V>0$, as shown in Fig.~\ref{fig:occupation}(b). In the non-interacting limit ($U=0$), any small $V$ suffices to trigger the charge transfer, and the {doublon response} persists as long as $n^{(-V)}$ continues to vary. As $V$ increases further, there is a critical $V_c^{metal \rightarrow band}$ at which the system crosses from this metallic phase to a band-like insulator regime, characterized by a maximum $n^{(-V)}$ and vanishing {doublon} response. 

For the interacting case with $m=0$, the system initially resides in a Mott-insulating phase. In particular, for {$U\gtrsim 1t$, due to the energy} cost of double occupancy, a minimum $V$ is required to trigger the charge transfer (Fig.~\ref{fig:occupation}(a)) and the {doublon} response (Fig.~\ref{fig:occupation}(b)). This minimum potential $V_c^{Mott{\rightarrow}metal}$ thus delineates the Mott insulator from the metallic phase. Upon further increase of $V$, $V>V_c^{metal \rightarrow band}$, the system reaches the band-like insulator regime. At $m=0$, a clear peak marks the metal-to-insulator transition for any interaction strength and also for low interactions ($U\lesssim3t$) at any magnetization. As $m$ increases, the metallic region shrinks ($V_c^{metal \rightarrow band}$ decreases) and the maximum $n^{(-V)}$ drops, since fewer opposite-spin electrons are available for {doublon formation}. Notably, $m$ does not affect $V_c^{Mott \rightarrow metal}$, confirming that the onset of metallicity stems purely from the competition between $U$ and $V$, while {doublon} formation in the band-insulating regime also depends on spin imbalance  \footnote{Notice that for $m\neq 0$, the band-like regime actually contains both band-like and ferromagnetic insulating states, due to the spin-imbalanced populations.}.

Fig.~\ref{fig:occupation} also reveals that the {doublon} mechanism $\partial \Bar{w_2}/\partial V$ exhibits a quasi-periodic oscillatory behavior. This reflects the interplay between $U$ and $V$: as $V$ increases, charge transfer and {doublon} formation are favored; on the other hand, the enhancement of charge/{doublons} also amplifies the relevance of $U$, which in turn diminishes the influence of $V$, causing then the oscillations. This interpretation is consistent with the fact that the oscillations essentially vanish at $U=0$ for any $m$. 

In Fig.~\ref{fig:occupation}(c) we analyze the behavior of the electronic hopping term,  $H_{t}$, as a function of $V$. We observe that within the metallic regime ($V_c^{Mott\rightarrow metal}<V<V_c^{metal\rightarrow band}$), $\langle H_{t} \rangle$ exhibits significant variation, reflecting the dynamic reorganization of charge induced by 
$V$. {This variation originates from tunneling processes associated with the hopping operators $c^{\dagger}_{i,\sigma}c_{j,\sigma}$, as reflected in the increase of $n^{(-V)}$ driven by electron motion toward the lower-potential region. Furthermore, increasing spin polarization $m$ narrows the range of $V$ for which variations in $\langle H_t \rangle$ occur, as also seen from the faster saturation of $n^{-V}$ in Fig.~\ref{fig:occupation}(a).} In contrast, the hopping contribution remains nearly constant for $V<V_c^{Mott{\rightarrow}metal}$ and $V>V_c^{metal{\rightarrow}band}$, indicating that the charge distribution remains unresponsive to $V$, which is consistent with the insulating phases.

\begin{figure}[t!]
    \centering
    \includegraphics[scale=0.6]{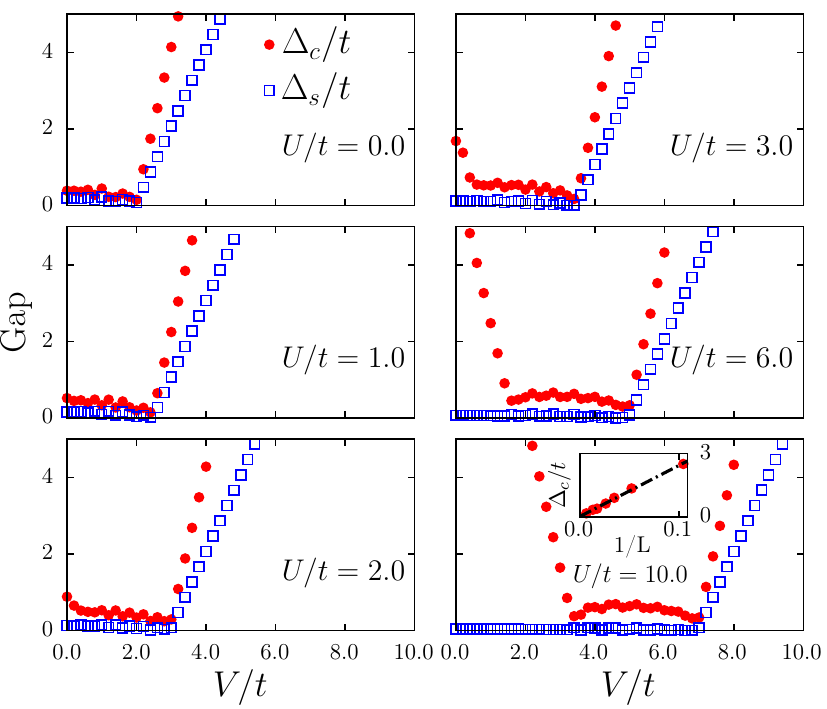}
    \caption{Charge and spin gaps, $\Delta_c$ and $\Delta_s$, respectively, as a function of the induced electric potential difference for $m=0$ and several interaction strengths $U$. The bottom panel includes an inset showing the linear decay in $L$ in the metallic region at $U/t=10$ e $V/t=4$.}
    \label{fig:gap}
\end{figure}

\subsection{Charge and spin gaps}

\begin{figure}[t!]
    \centering
    \includegraphics[scale=0.55]{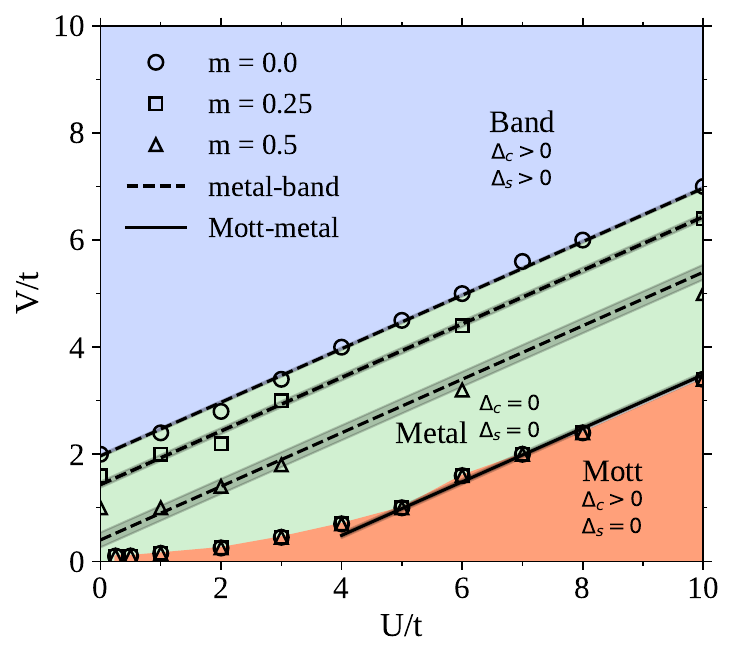}
    \caption{Phase diagram for different values of $m$, delimitating the Mott-insulator, metallic, and band-insulator phases. Symbols are numerical data, while lines represent their linear fits (shaded areas are their standard deviations): {$V_c^{Mott\rightarrow metal}=U/2-1.51t$ for all $m$ (within $U > 3t$)}, {and for any $U$} $V_c^{metal\rightarrow band}=U/2+b(m)t$, with $b(m)=\{1.97;~1.43;~0.40\}$ for $m=\{0;~ 0.25;~0.5\}$, respectively.}   
\label{diagramUV}
\end{figure}

The charge ($\Delta_{c}$) and spin ($\Delta_{s}$) gaps as a function of $V$ are presented in Fig.~\ref{fig:gap} for several interactions at $m=0$. We find that $\Delta_{c}$ is finite for the two insulating regimes, tending to zero exclusively within the range $V_c^{Mott\rightarrow metal} < V < V_c^{metal\rightarrow band}$, {decaying linearly with} $L$ (see the inset), which confirms its metallic character {in the low-energy sector}. {This behavior contrasts with previous studies~\cite{naoto_95_2005,dagotto_74_2006} where the electric potential is determined self-consistently, which smooths the step potential and confines the metallic response to a finite-width region around the interface. In our case, the linear scaling with $1/L$, observed up to $L=64$, suggests that the metallic regime may extend beyond the interface region. However, to determine its true spatial extent, simulations on larger chains are necessary. Since the correlation length scales as $1/\Delta_C$, it can easily exceed the system size when the gap becomes small, potentially masking the actual boundaries of the metallic regime.} Moreover, recent results have shown that interfaces can also display insulating, rather than metallic, behavior \cite{nair_109_2024}, highlighting that the interface response is not universal. For $V>V_c^{metal\rightarrow band}$, the spin gap is also finite $\Delta_s\neq 0$, which is consistent with a band-like insulator \cite{PhysRevA.72.013604,PhysRevA.70.031603}.
Interestingly, for \( V<V_c^{metal\rightarrow band} \) the system exhibits gapless spin excitations ($\Delta_s=0$), thus characterizing a state resembling a spin-density wave state \cite{EjimaEHM2007}.
{This indicates that, although the spin sector is gapless, the observables shown in Fig.~\ref{fig:occupation} remain essentially unchanged as $m$ increases, suggesting that spin excitations do not significantly impact these specific quantities within our numerical resolution.}

Hence the three regimes are characterized by the gaps as follows: in the Mott-insulator phase $\Delta_c >0 $ and $\Delta_s =0$; at the metallic regime $\Delta_c =0$ and $\Delta_s=0$; and at the band-like insulator phase $\Delta_c >0 $ and $\Delta_s >0 $. Fig.~\ref{diagramUV} shows the resulting phase diagram in the $U - V$ plane for distinct values of magnetization. {For $m=0$, the transition lines are determined from the finite-size scaling of the charge gap. For $m>0$, where the charge gap is not well defined, the metallic boundaries are instead determined from the behavior of the other quantities --- the electric-potential dependence of the kinetic energy and the oscillations in local doublon response and entanglement --- which exhibit the same characteristic trends as in the gapless metallic phase at $m=0$. In this sense, the identification of metallic regions at finite magnetization follows a qualitative criterion guided by the $m=0$ case.} According to the above discussion, the transition line $V_{c}^{Mott\rightarrow metal}$ remains unchanged for $m>0$, while the boundary $V_{c}^{metal\rightarrow band}$ shifts, progressively narrowing the metallic region.

\subsection{Entanglement}

\begin{figure}[t!]
    \centering
    \includegraphics[scale=0.7]{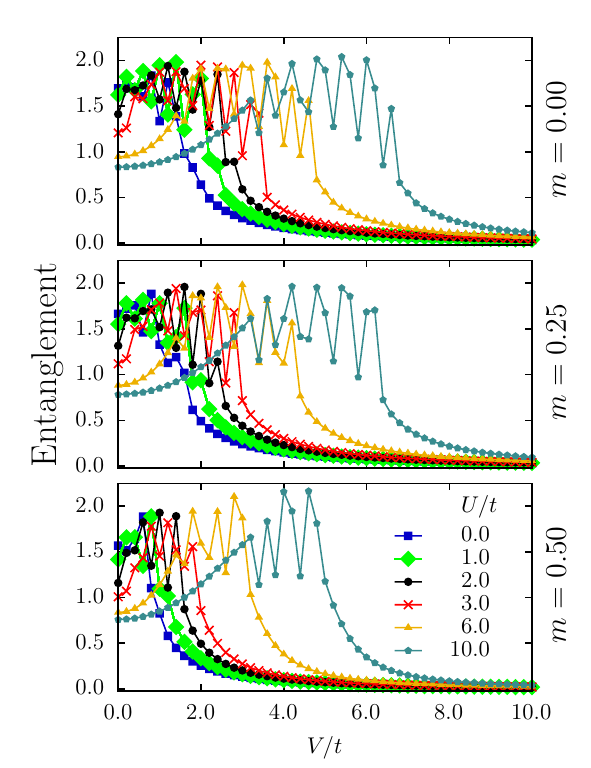}
    \caption{Entanglement between the two halves of the chain under an applied electric potential difference, for several interaction strengths $U$ and magnetizations $m$. }
    \label{fig:entanglement}
\end{figure}

Fig.~\ref{fig:entanglement} presents the entanglement $S$ as a function of the electric {potential} for several $U$ and $m$. Clearly $S$ exhibits a quasi-periodic oscillatory behavior exclusively at the metallic regime, which is narrowed for $m>0$, similar to the {doublon} response (Fig.~\ref{fig:occupation}). The entanglement oscillations arise from enhanced charge and spin fluctuations at the interface, as the bipartition of $S$ is located precisely where the electric potential abruptly changes. Remarkably, $m$ practically has no effect on the maximum entanglement $S_{max}\approx 2$, indicating a robust upper bound within the metallic phase. {Consistent with the 1D area law~\cite{RevModPhys.82.277} --- which implies that entanglement counts a finite, non-extensive set of effective states across the interface --- the persistence of $S_{\max}$ under spin polarization is also a robust signature that the metallic regime survives for $m>0$, since insulating phases suppress interfacial correlations and therefore yield lower entanglement.}

For the Mott and band-like insulators, entanglement reveals the localized nature of electronic states through its asymptotic behavior. For $V \gg U$, entanglement approaches zero, signaling the suppression of quantum correlations {across the interface} between the two halves of the chain. This is expected, as in this limit the system effectively behaves as if an infinite potential barrier forms at the interface, decoupling the halves. On the other hand, for $U \gg V$, since charge fluctuations are suppressed, finite entanglement originates solely from spin degrees of freedom, due to the superexchange interaction~\cite{essler2005one,giamarchi2003quantum}, which mediates spin correlations across the interface {and underlies the spin-density wave order}.

\section{Conclusion}\label{sec13}

In summary, we investigated quantum phase transitions induced by a {step-like} electric potential $V$ in a 1D half-filled Hubbard model. The interplay between electronic interactions and the potential $V$ defines distinct regimes in the system: Mott-insulator, metallic, and band-insulator phases. The metallic phase, marked by the closing of both charge and spin gaps, is accompanied by quasi-periodic oscillatory behavior in {local doublon response} and entanglement, due to quantum fluctuations enhanced at the interface. Although the metallic regime persists across different magnetizations, higher spin polarization reduces its width in the phase diagram, Fig.~\ref{diagramUV}. Future work could aim to {explore how} smoother interfaces {affect the phase diagram. For instance, a linear Wannier–Stark potential \cite{nair_109_2024} --- which may be regarded as a smoother version of our step potential --- has been shown to yield insulating domains at the interfaces, suggesting that smoothing may lead to qualitatively different scenarios. Other directions include extending} the analysis to finite temperature and higher dimensionality, {as well as proposing} experimental setups for realizations in ultracold atoms or 1D materials.

%\backmatter

%\bmhead{Supplementary information}

%If your article has accompanying supplementary file/s please state so here. 

%Authors reporting data from electrophoretic gels and blots should supply the full unprocessed scans for key as part of their Supplementary information. This may be requested by the editorial team/s if it is missing.

%Please refer to Journal-level guidance for any specific requirements.

\bmhead{Acknowledgements}

This work was supported by São Paulo Research Foundation Fapesp ($2021/06744$-$8$, $2024/10789$-$5$, $2023/00510$-$0$) and the National Council of
Technological and Scientific Development CNPq ($403890/2021$-$7$, $140854/2021$-$5$).

\section*{Declarations} 

\section*{Data Availability Statement} Data can be provided on reasonable request to the corresponding author.

\section*{Author Contribution Statement} DA and RMD carried out the numerical calculations. All authors contributed to the analysis of the results and the writing of the manuscript.

\end{document}